
\input harvmac
\def\H{{\cal H}}
\def\O{{\cal O}}
\def\mod{{\rm mod}}
\def\lp{\lambda^\prime}
\def\lpp{\lambda^{\prime\prime}}
\def\kp{\kappa^\prime}
\def\kpp{\kappa^{\prime\prime}}

\def\rp{\rho^\prime}
\def\rpp{\rho^{\prime\prime}}

\Title{hep-ph/9408315, WIS-94/34/Aug-PH}
{\vbox{\centerline{Implications of Horizontal Symmetries on}
 \centerline{Baryon Number Violation in Supersymmetric Models}}}
\bigskip
\centerline{Valerie Ben-Hamo and Yosef Nir}
\smallskip
\centerline{\it Department of Particle Physics}
\centerline{\it Weizmann Institute of Science, Rehovot 76100, Israel}
\bigskip
\baselineskip 18pt

\noindent
The smallness of the quark and lepton parameters and the
hierarchy between them could be the result of selection rules due to
a horizontal symmetry broken by a small parameter.
The same selection rules apply to baryon number violating terms.
Consequently, the problem of baryon number violation in Supersymmetry
may be solved naturally, without invoking any especially-designed
extra symmetry. This mechanism is efficient enough even for low-scale
flavor physics. Proton decay is likely to be dominated by the modes
$K^+\bar\nu_i$ or $K^0\mu^+(e^+)$, and may proceed at observable rates.

\Date{8/94}

The smallness and the hierarchy in the quark and lepton
parameters may be related to a
horizontal symmetry $\H$ that acts on the fermions.
Such a horizontal symmetry may be responsible for the hierarchy
if it is explicitly broken by an operator in the Lagrangian
whose coefficient is a small parameter $\lambda$.
(Numerically, we expect $\lambda\sim0.2$ to explain the
Cabibbo angle.)
The transformation laws of $\lambda$ under $\H$ control the order in
perturbation theory of the various elements in the fermion
mass matrices (``selection rules") and, consequently, some parameters
depend on powers of $\lambda$ higher than others,
namely a hierarchy can be generated.

Supersymmetric models allow, in general, baryon number and lepton
number violation
\ref\DiGe{S. Dimopoulos and H. Georgi, Nucl. Phys. B193 (1981) 150;
S. Weinberg, Phys. Rev. D26 (1982) 287;
N. Sakai and T. Yanagida, Nucl. Phys. B197 (1982) 533;
J. Ellis, J. Hagelin, D. Nanopoulos and K. Tamvakis,
 Phys. Lett. B124 (1983) 484;
P. Nath, A.H. Chamseddine and R. Arnowitt, Phys. Rev. D32 (1985) 2348.}.
If the relevant couplings are of $\O(1)$,
then proton decay and neutron-antineutron oscillations are
predicted at unacceptably fast rates. One may try to solve this
problem by invoking additional symmetries designed particularly
to forbid the problematic terms
\ref\DRW{See {\it e.g.} S. Dimopoulos, S. Raby and F. Wilczek,
 Phys. Lett. B112 (1982) 133;
A. Font, L. Ib\'a\~nez and F. Quevedo, Phys. Lett. B228 (1989) 79;
L. Ib\'a\~nez and G.G. Ross, Nucl. Phys. B368 (1992) 3;
I. Hinchliffe and T. Kaeding, Phys. Rev. D47 (1993) 279;
P.L. White, Nucl. Phys. B403 (1993) 141.}.
(The problem may also be solved with specific gauge symmetries
\ref\ELN{See {\it e.g.} J. Ellis, J. Lopes and D.V. Nanopoulos,
 Phys. Lett. B252 (1990) 53;
G. Leontaris and T. Tamvakis, Phys. Lett. B260 (1991) 333;
A.E. Faraggi, IASSNS-HEP-94/18, hep-ph/9403312.}.)
However, the selection rules
due to the horizontal symmetry apply to baryon violating terms
in the Lagrangian as well and are likely to suppress them.
It is the purpose of this work to see whether this mechanism
can naturally solve the problem of baryon number violation
in supresymmetric theories.
(A similar question has been recently taken in ref.
\ref\MuKa{H. Murayama and D.B. Kaplan, NSF-ITP-94-69, hep-ph/9406423.}.)

We work in the framework of supersymmetric Abelian horizontal
symmetries that has been recently investigated in refs.
\ref\lnsa{M. Leurer, Y. Nir and N. Seiberg, Nucl. Phys. B398 (1993) 319.}
\ref\lnsb{Y. Nir and N. Seiberg, Phys. Lett. B309 (1993) 337.}
\ref\lnsc{M. Leurer, Y. Nir and N. Seiberg,
 Nucl. Phys. B420 (1994) 468.}.
 (For recent related work, see
\ref\DLK{M. Dine, R. Leigh and A. Kagan, Phys. Rev. D48 (1993) 2214;
P. Pouliot and N. Seiberg, Phys. Lett. B318 (1993) 169;
D.B. Kaplan and M. Schmaltz, Phys. Rev. D49 (1994) 3741;
L. Ib\'a\~nez and G.G. Ross, Phys. Lett. B332 (1994) 100.}.)
We assume that the low energy spectrum
consists of the fields of the minimal supersymmetric Standard Model with
the following $SU(3)_C\times SU(2)_L\times U(1)_Y$ quantum numbers:
\eqn\LEspectrum{\eqalign{
Q_i&(3,2)_{1/6},\ \bar u_i(\bar3,1)_{-2/3},\ \bar d_i(\bar3,1)_{1/3},\
L_i(1,2)_{-1/2},\ \bar\ell_i(1,1)_1,\cr
\phi_u&(1,2)_{1/2},\ \phi_d(1,2)_{-1/2},\cr}}
where $i=1,2,3$ is a generation index. Each of these fields
carries a charge under an Abelian horizontal symmetry $\H$.
For most of our discussion, it makes no difference whether $\H$
is local or global, continuous or discrete. $\H$ is explicitly
broken by a small parameter $\lambda$ to which we attribute charge --1.
Then, the following selection rules apply:
\item{a.} Terms in the superpotential that carry charge $n\geq0$
under $\H$ are suppressed by $\O(\lambda^n)$, while those with
$n<0$ are forbidden due to the holomorphy of the superpotential.
(If $\H$ is a discrete $Z_N$, the suppression is
by $\O(\lambda^{n(\mod N)})$.)
\item{b.} Terms in the K\"ahler potential that carry charge $n$
under $\H$ are suppressed by $\O(\lambda^{|n|})$ (or $\O(\lambda^{
n(\mod N)})$ for $\H=Z_N$).
\par
The resulting fermion mass matrices should give the following
orders of magnitude for the various physical parameters
(we assume $\tan\beta\sim1$):
\eqn\order{\eqalign{
1\sim&\ m_t/\vev{\phi_u},\cr  \lambda\sim&\ |V_{us}|,\cr
\lambda^2\sim&\ |V_{cb}|,\ m_d/m_s,\ m_s/m_b,\ m_b/m_t,\ m_\mu/m_\tau,\cr
\lambda^3\sim&\ |V_{ub}|,\ m_u/m_c,\ m_c/m_t,\ m_e/m_\mu,\ m_\tau/m_t
.\cr}}
Some of these order of magnitude estimates are ambiguous,
{\it e.g.} $m_b/m_t\sim\lambda^3-\lambda^2$.
Furthermore, the estimates depend on the scale. Also, it could be that
$m_b/m_t$ is explained by large $\tan\beta$ and that the bare $m_u$
vanishes. None of these points changes the principles of our mechanism,
so we will only study models that satisfy the hierarchy as given in
\order. It is straightforward to change the specific details of our
examples to take account of other options.

The following terms are relevant to proton decay and neutron-antineutron
oscillations (gauge indices are suppressed below):
\item{(a)} Dimension-4 terms from the superpotential:
\eqn\fourF{\left[\lp_{ijk}L_iQ_j\bar d_k+
\lpp_{ijk}\bar u_i\bar d_j\bar d_k\right]_F.}
\item{(b)} Dimension-5 terms from the superpotential:
\eqn\fiveF{{1\over M}\left[\kp_{ijkl}Q_iQ_jQ_kL_l+
\kpp_{ijkl}\bar u_i\bar u_j\bar d_k\bar\ell_l
\right]_F.
}
\item{(c)} Dimension-6 terms from the K\"ahler potential:
\eqn\sixD{{1\over M^2}
\left[\rp_{ijkl}\bar u^\dagger_i\bar d^\dagger_j Q_kL_l+
\rpp_{ijkl}Q_iQ_j\bar u_k^\dagger\bar\ell_l^\dagger\right]_D.}
\par
There are also dimension-5 terms from the K\"ahler potential,
as well as other terms of each of the three types given above. However,
within our framework, these additional $B$ or $L$ violating terms
never give the dominant contribution to the processes that we study.

When SUSY particles are integrated out,
combinations of the above operators give effective $B$- and
$L$-violating four-fermi operators of the general type
$\eta_{\rm eff}qqq\ell$ (where generation, gauge and Lorentz indices
are suppressed) that lead to proton decay. The upper bounds on proton
decay rates
\ref\PDG{K. Hikasa {\it et al.}, Phys. Rev. D45 (1992) VIII.3.}\
require $\eta_{\rm eff}\leq10^{-32}\ GeV^{-2}$.
This can be translated to the following bounds on the $\lambda$,
$\kappa$ and $\rho$ couplings:
\eqn\lbound{{\lp_{ijk}\lpp_{11k}\over M_{\rm SUSY}^2}\leq10^{-32}\
GeV^{-2}}
with $i=1,2,3,\ j=1,2,\ k=2,3$.
\eqn\kbound{\eqalign{
{\kp_{112i}\over M_{\rm SUSY}M}\leq&\ 10^{-29}\ GeV^{-2},\cr
{\kpp_{1jkl}(K^u_{RR})_{1j}\over M_{\rm SUSY}M}\leq&\ 10^{-30}\
GeV^{-2},\cr}}
with $i=1,2,3,\ j=2,3,\ k,l=1,2$. (We have taken ${\alpha_2\over4\pi}
{m_{\tilde w}\over m_{\tilde q}}\sim10^{-3}$ and ${\alpha_3\over4\pi}
{m_{\tilde g}\over m_{\tilde q}}\sim10^{-2}$. The $K^q_{MN}$ mixing
matrices for gaugino couplings are defined in \lnsb.)
\eqn\rbound{\eqalign{
{\rp_{1ijk}\over M^2}\leq&\ 10^{-32}\ GeV^{-2},\cr
{\rpp_{1l1m}\over M^2}\leq&\ 10^{-32}\ GeV^{-2},\cr}}
with $(i,j)=(1,2),(2,1),(1,1),\ k=1,2,3,\ l,m=1,2$.
If, for example, we take $M\sim{M_p\over\sqrt{8\pi}}=2.4\times10^{18}$
$GeV$ and $M_{\rm SUSY}\sim10^3\ GeV$, then \lbound\ and \kbound,
expressed in powers of $\lambda=0.2$, become
\eqn\lboundl{\lp_{ijk}\lpp_{11k}\leq\lambda^{37},}
\eqn\kboundl{\kp_{112i}\leq\lambda^{11},
\ \ \kpp_{1jkl}(K^u_{RR})_{1j}\leq\lambda^{12},}
while \rbound\ allows $\rho=\O(1)$. If we assume squark
degeneracy, $(K^u_{RR})_{12}=0$ and there is no constraint on $\kpp$.

{\it In principle, it is always possible to find a horizontal
symmetry $\H$ that gives the required suppression of
the $B$ violating couplings.} To see that, note that the
Yukawa terms have an accidental symmetry $U(1)_B\times U(1)_L
\times U(1)_X$ (under $U(1)_X$, $\phi_d$ carries charge --1,
$\bar d_i$ and $\bar\ell_i$
carry charge +1, and all other fields are neutral).
Therefore, requiring \order\ fixes the horizontal charges only
up to arbitrary shifts by $\alpha B+\beta L+\gamma X$.
In ref. \lnsa, which was concerned with fermion mass matrices only,
this symmetry of the Yukawa terms (together with the gauge $U(1)_Y$)
was used to put the $\H$-charges of $\phi_u$, $\phi_d$, $Q_3$ and
$L_3$ to zero. However, the $B$ and $L$ violating terms that we
investigate here are {\it not} invariant under this symmetry.
Thus we can always find a symmetry
\eqn\tildeH{\tilde\H\subset\H\times U(1)_B\times U(1)_L\times U(1)_X}
that is {\it isomorphic} to $\H$ but will give an arbitrarily strong
suppression
of the $\Delta B\neq0$ terms. The symmetry $\tilde\H$ will, of course,
dictate precisely the same mass matrices as $\H$, but at the same time
it will solve the baryon-number violation problem of Supersymmetry
without invoking any additional ad-hoc symmetry.
(Actually, as long as the only input from the lepton sector are
the charged lepton masses, we have the freedom of $U(1)_e\times
U(1)_\mu\times U(1)_\tau$ rather than just $U(1)_L$.)

The only potential drawback in this mechanism is that the
required $\tilde\H$ charges may turn out to be very large, in which case
the model becomes
unnatural and is unlikely to be realized in nature. To make this
point clear, $\tilde\H=\H+100B$ (where $\H$ is the horizontal symmetry
under which $\phi_u$, $\phi_d$, $Q_3$ and $L_3$ are neutral)
would certainly satisfy all the
constraints in \lboundl\ and \kboundl. But when the various quark and
lepton supermultiplets carry $\tilde\H$ charges of $\O(100)$ (in
units of the $\tilde\H$-charge of $\lambda$), the model is not very
plausible. The real test of this mechanism is then whether it can solve
the baryon violation problem with reasonable charges, say $\leq10$.
Below we give three examples to demonstrate that, indeed, rather simple
horizontal symmetries with reasonable charges to all fields
can suppress baryon number violation to an acceptable degree.

First, we take the ``master model" of ref. \lnsc. The horizontal
symmetry is $\H=U(1)_H$ with a small breaking parameter $\lambda$
carrying $H=-1$. Consider the following set of charge assignments:
\eqn\mastercharge{\matrix{Q_1&Q_2&Q_3&&\bar d_1&\bar d_2&\bar d_3&&
\bar u_1&\bar u_2&\bar u_3\cr
(3)&(2)&(0)&&(9)&(8)&(8)&&(3)&(1)&(0)\cr
L_1&L_2&L_3&&\bar\ell_1&\bar\ell_2&\bar\ell_3&&\phi_u&\phi_d&\cr
(7)&(7)&(7)&&(7)&(4)&(2)&&(0)&(-6).&\cr}}
It leads to the following fermion mass matrices:
\eqn\mastermasses{
M^d\sim\vev{\phi_d}\pmatrix{\lambda^6&\lambda^5&\lambda^5\cr
\lambda^5&\lambda^4&\lambda^4\cr \lambda^3&\lambda^2&\lambda^2\cr},\ \ \
M^u\sim\vev{\phi_u}\pmatrix{\lambda^6&\lambda^4&\lambda^3\cr
\lambda^5&\lambda^3&\lambda^2\cr
\lambda^3&\lambda&1\cr},\ \ \
M^\ell\sim\vev{\phi_d}\pmatrix{\lambda^8&\lambda^5&\lambda^3\cr
\lambda^8&\lambda^5&\lambda^3\cr \lambda^8&\lambda^5&\lambda^3\cr}.}
This gives the required order of magnitude estimates \order.
At the same time, \mastercharge\ leads to
\eqn\lmaster{\lp_{i23}\sim\lambda^{17},\ \ \ \lpp_{113}\sim\lambda^{20},}
\eqn\kmaster{\kp_{112j}\sim\lambda^{15},}
which satisfy \lboundl\ and \kboundl. (We assume here squark
degeneracy so that $\kpp$ poses no problem, and $M={M_p\over\sqrt{8\pi}}$
so that $\rp$ and $\rpp$ give negligible contributions.)
The leading proton decay mode, due to $\lp_{i2j}\lpp_{11j}$, is
\eqn\mastermode{p\rightarrow K^+\bar\nu_i.}

One of the main purposes of refs. \lnsa\ and \lnsc\ was to check whether
the flavor physics scale could be at low enough energy to be
directly accessible in future experiments. The answer was that
this is possible though not very likely.
Now the following question arises: if we require
that the horizontal symmetry solves the baryon number violation
problem in the manner described above, is it still possible to have
$M$ as low as $\sim10^5\ GeV$? (In the examples of a full high-energy
theory in refs. \lnsa\lnsc, based on the model of ref.
\ref\FrNi{C.D. Froggatt and H.B. Nielsen, Nucl. Phys. B147 (1979) 277.},
$M$ is the mass scale for heavy fermions in vector representations.)
Assuming squark degeneracy, that would require
\eqn\lboundlow{\lp_{ijk}\lpp_{11k}\leq\lambda^{37},}
\eqn\kboundlow{\kp_{112i}\leq\lambda^{30},}
\eqn\rboundlow{\rp_{112i},\ \rpp_{111j}\leq\lambda^{31}.}
Had we found that this is possible only with very high $\H$-charges,
we should have concluded that the ideas of
low-energy flavor physics and of $B$-violation suppressed by $\H$
are mutually exclusive.

The models where a low scale could be consistent with FCNC and Landau
poles constraints employed $\H=U(1)_{H_1}\times U(1)_{H_2}$. There are
two small breaking parameters: $\lambda_1\sim\lambda^2$ and $\lambda_2
\sim\lambda^3$ with $(H_1,H_2)$ charges $(-1,0)$
and $(0,-1)$, respectively.
Our second example is then a model with this horizontal
symmetry and the following charge assignments:
\eqn\lowcharge{\matrix{Q_1&Q_2&Q_3&&\bar d_1&\bar d_2&\bar d_3&&
\bar u_1&\bar u_2&\bar u_3\cr
(5,3)&(6,2)&(5,2)&&(3,2)&(1,3)&(1,3)&&(-5,-1)&(-6,-1)&(-5,-2)\cr
L_1&L_2&L_3&&\bar\ell_1&\bar\ell_2&\bar\ell_3&&\phi_u&\phi_d&\cr
(1,4)&(2,3)&(5,6)&&(5,3)&(4,3)&(0,0)&&(0,0)&(-5,-5).&\cr}}
It leads to the following fermion mass matrices:
\eqn\lowmasses{\eqalign{
M^d&\sim\vev{\phi_d}\pmatrix{
\lambda_1^3&\lambda_1\lambda_2&\lambda_1\lambda_2\cr
0&\lambda_1^2&\lambda_1^2\cr 0&\lambda_1&\lambda_1\cr},\ \ \
M^u\sim\vev{\phi_u}\pmatrix{
\lambda_2^2&0&\lambda_2\cr \lambda_1\lambda_2&\lambda_2&\lambda_1\cr
\lambda_2&0&1\cr},\cr
M^\ell&\sim\vev{\phi_d}\pmatrix{
\lambda_1\lambda_2^2&\lambda_2^2&0\cr
\lambda_1^2\lambda_2&\lambda_1\lambda_2&0\cr
\lambda_1^5\lambda_2^4&\lambda_1^4\lambda_2^4&\lambda_2\cr}.\cr}}
This gives the order of magnitude estimates \order.
At the same time, \lowcharge\ leads to
\eqn\llow{\lp_{223}\sim\lambda_1^9\lambda_2^8,\ \ \ \lpp_{113}=0,}
\eqn\klow{\kp_{1122}\sim\lambda_1^{18}\lambda_2^{11},}
\eqn\rlow{\rp_{1122}\sim\lambda_1^{10}\lambda_2^4,\ \ \
\rpp_{1211}\sim\lambda_1^{11}\lambda_2^3,}
which satisfy \lboundlow, \kboundlow\  and \rboundlow.
The vanishing of $\lpp_{113}$ comes from holomorphy and would
be lifted if the symmetry is discrete. (We, again, assume here squark
degeneracy so that $\kpp$ poses no problem.)
The leading proton decay mode, due to $\rpp_{1211}$, is
\eqn\lowmode{p\rightarrow K^0\bar e^+.}
As emphasized in ref. \MuKa, this mode does not typically arise in
SUSY GUT models and is likely to signify flavor physics of the
type described in this work.

As our third example, we take the quark-squark alignment models
of ref. \lnsc. This class of models gives a suppression of
FCNC from supersymmetric diagrams by forcing the quark mass
matrices and squark mass-squared matrices to be simultaneously
approximately diagonal \lnsb. No squark degeneracy is needed.
The horizontal symmetry is $\H=U(1)_{H_1}\times U(1)_{H_2}$.
There are two small breaking parameters, $\lambda_1\sim\lambda$
and $\lambda_2\sim\lambda^2$ with $(H_1,H_2)$ charges $(-1,0)$
and $(0,-1)$, respectively. Consider the following charge assignments:
\eqn\qsacharge{\matrix{Q_1&Q_2&Q_3&&\bar d_1&\bar d_2&\bar d_3&&
\bar u_1&\bar u_2&\bar u_3\cr
(1,1)&(-2,2)&(-2,1)&&(1,4)&(6,1)&(2,3)&&(1,1)&(3,-1)&(2,-1)\cr
L_1&L_2&L_3&&\bar\ell_1&\bar\ell_2&\bar\ell_3&&\phi_u&\phi_d&\cr
(5,1)&(-1,4)&(1,3)&&(-3,5)&(2,1)&(0,1)&&(0,0)&(0,-3).&\cr}}
It leads to the following fermion mass matrices:
\eqn\qsamasses{\eqalign{
M^d&\sim\vev{\phi_d}\pmatrix{
\lambda_1^2\lambda_2^2&0&\lambda_1^3\lambda_2\cr
0&\lambda_1^4&\lambda_2^2\cr 0&0&\lambda_2\cr},\ \ \
M^u\sim\vev{\phi_u}\pmatrix{
\lambda_1^2\lambda_2^2&\lambda_1^4&\lambda_1^3\cr
0&\lambda_1\lambda_2&\lambda_2\cr 0&\lambda_1&1\cr},\cr
M^\ell&\sim\vev{\phi_d}\pmatrix{
\lambda_1^2\lambda_2^3&0&0\cr 0&\lambda_1\lambda_2^2&0\cr
0&\lambda_1^3\lambda_2&\lambda_1\lambda_2\cr}.\cr}}
Note, in particular, the various zero entries in $M^d$ which suppress
the SUSY contribution to $K-\bar K$ mixing. The charge assignments
\qsacharge\ lead to
\eqn\lqsa{\lp_{323}\sim\lambda_1\lambda_2^8,\ \ \
\lpp_{113}\sim\lambda_1^4\lambda_2^8,}
\eqn\kqsa{\kp_{1123}\sim\lambda_1\lambda_2^7,\ \ \
\kpp_{1322}\sim\lambda_1^{11}\lambda_2^2,\ \ \
(K^u_{RR})_{13}\sim\lambda_1\lambda_2^2,}
which satisfy \lboundl\ and \kboundl. Note the need to consider
$\kpp_{1ijk}(K^u_{RR})_{1i}$ as squarks are not necessarily degenerate.
(We, again, take $M={M_p\over\sqrt{8\pi}}$
so that $\rp$ and $\rpp$ give negligible contributions.)
The leading decay mode, due to $\lp_{i23}\lpp_{113}$, is
\eqn\qsamode{p\rightarrow K^+\bar\nu_i.}

The models presented above assume that there is no additional
symmetry that forbids the $B$ and $L$ violating terms.
It could be that there exists a discrete
$R$-parity, $R_p$, which forbids the $\lambda$ couplings
of eq. \fourF,
as well as dimension-5 D-terms.
In this case, the only dangerous terms are the $\kappa$ couplings
of eq. \fiveF\ (and the $\rho$ couplings of eq. \sixD,
if the scale $M$ is below $10^{16}\ GeV$). This scenario was
recently investigated in an interesting paper by Murayama and Kaplan
\MuKa. In our framework, this scenario makes the constraints
much easier to satisfy. With degenerate squarks and $M={M_p\over
\sqrt{8\pi}}$ (as assumed in \MuKa),
the only important constraint is $\kp_{1jkl}\leq\lambda^{11}$
which is easily satisfied with small $\H$ charges, {\it e.g.}
\eqn\Rprcharge{\matrix{Q_1&Q_2&Q_3&&\bar d_1&\bar d_2&\bar d_3&&
\bar u_1&\bar u_2&\bar u_3\cr
(3)&(2)&(0)&&(1)&(0)&(0)&&(3)&(1)&(0)\cr
L_1&L_2&L_3&&\bar\ell_1&\bar\ell_2&\bar\ell_3&&\phi_u&\phi_d&\cr
(3)&(3)&(3)&&(3)&(0)&(-2)&&(0)&(2).&\cr}}

We have also investigated the constraints from $n-\bar n$ oscillations.
When SUSY particles are integrated out,
combinations of the operators under study give effective
$B$-violating six-fermi operators of the general type
$\sigma_{\rm eff}qqqqqq$ (where generation, gauge and Lorentz
indices are suppressed) that lead to neutron-antineutron oscillations.
The upper bound on the rate of $n-\bar n$ oscillations
\ref\PDGn{K. Hikasa {\it et al.}, Phys. Rev. D45 (1992) VIII.8.}\
requires $\sigma_{\rm eff}\leq10^{-27}\ GeV^{-5}$.
This gives, for example, the following bounds on $\lpp$
\ref\Zwi{F. Zwirner, Phys. Lett. 132B (1983) 103.}:
\eqn\lboundn{\eqalign{{\lpp_{112}\lpp_{113}\lpp_{323}\lpp_{312}
\over M_{\rm SUSY}^5}\leq&\ 10^{-27}\ GeV^{-5},\cr
{\lpp_{112}\lpp_{113}(K^d_{RR})_{12}(K^d_{RR})_{13}
\over M_{\rm SUSY}^5}\leq&\ 10^{-27}\ GeV^{-5}.\cr}}
We find, however, that these bounds are always satisfied once
those from proton decay are.

To summarize the main conclusions of this work:

(a) Abelian horizontal symmetries that explain the
smallness and hierarchy in the quark and lepton sector parameters,
may at the same time suppress baryon number violating couplings
to an acceptable degree. There is no need to invoke extra symmetries
for the sole purpose of forbidding the $\Delta B\neq0$ terms.

(b) For models of horizontal symmetries where a phenomenologically
interesting scale for flavor physics is consistent with
FCNC and Landau poles constraints, the constraints from proton
decay and $n-\bar n$ oscillations can still be satisfied with
the same low scale.

(c) Operators that do not contribute to proton decay when squarks
are degenerate do contribute in models of quark--squark alignment but,
again, can be satisfactorily suppressed by the horizontal symmetry.

(d) If the suppression of proton decay is due to a horizontal
symmetry, then the leading decay modes are to final kaons, {\it i.e.}\
$K^+\bar\nu_i$ or $K^0\mu^+(e^+)$. In the absence of information
about neutrino masses and mixings, no analogous statement can be made
about the final leptons.

(e) Unlike $R_p$ which forbids certain terms, the horizontal symmetry
could either forbid (F terms) or suppress them. Furthermore,
the possibility of Supersymmetry without $R$-parity leads to many other
interesting phenomenological consequences
\ref\HaSu{See {\it e.g.}
L.J. Hall and M. Suzuki, Nucl. Phys. B231 (1984) 419;
R. Mohapatra, Phys. Rev. D34 (1986) 3457;
R. Barbieri and A. Masiero, Nucl. Phys. B267 (1986) 679;
S. Dimopoulos and L. Hall, Phys. Lett. B196 (1987) 135;
V. Barger, G.F. Giudice and T. Han, Phys. Rev. D40 (1989) 2987;
R. Barbieri, D.E. Brahm, L.J. Hall and S.D.H. Hsu,
 Phys. Lett. B238 (1990) 86;
H. Dreiner and G.G. Ross, Nucl. Phys. B365 (1991) 597;
S. Lola and J. McCurry, Nucl. Phys. B381 (1992) 559.}.

The fact that horizontal symmetries that are invoked to explain
the hierarchy in fermion parameters may solve many
other problems -- FCNC in Supersymmetry \lnsb, the $\mu$-problem \lnsc,
hierarchy of symmetry breaking scales \lnsc, the strong CP problem
\ref\bns{T. Banks, Y. Nir and N. Seiberg, RU-94-24, hep-ph/9403203.},
FCNC due to light leptoquarks
\ref\BaLe{E. Baver and M. Leurer, WIS-94/27/Jul-PH, hep-ph/9407324.},
and baryon number violation in Supersymmetry as described in \MuKa\
and in this work -- makes this extension of the Standard Model a very
attractive one.

\centerline{\bf Acknowledgements}
We thank Nati Seiberg for useful discussions.
Y.N. is an incumbent of the Ruth E. Recu Career Development chair,
and is supported in part by the Israel Commission for Basic Research,
by the United States--Israel Binational Science Foundation (BSF),
and by the Minerva Foundation.

\listrefs

\end